\documentclass[aps,pra,twocolumn,showpacs,superscriptaddress,nofootinbib,10pt]{revtex4-1}
\usepackage{amsmath,amssymb,graphicx}

\usepackage{dcolumn}
\usepackage{braket}
\usepackage{amssymb}
\usepackage{amsmath}
\usepackage{latexsym}
\usepackage{mathrsfs}
\usepackage{arydshln}
\usepackage[sans]{dsfont}
\usepackage{epsfig}
\usepackage{epstopdf}

\newcommand{\bitem}{\begin{itemize}}
\newcommand{\fitem}{\end{itemize}}
\newcommand{\beq}{\begin{equation}}
\newcommand{\eeq}{\end{equation}}
\newcommand{\beqa}{\begin{eqnarray}}
\newcommand{\eeqa}{\end{eqnarray}}

\begin{document}

\title{Spatial and temporal localization of light in two dimensions}

\author{C. E. M\'aximo} \affiliation{Instituto de F\'{i}sica de S\~ao Carlos, Universidade de S\~ao Paulo, 13560-970 S\~ao Carlos, SP, Brazil}
\author{N. Piovella} \affiliation{Dipartimento di Fisica, Universit\`a degli Studi di Milano, Via Celoria 16, Milano I-20133, Italy}
\author{Ph. W. Courteille} \affiliation{Instituto de F\'{i}sica de S\~ao Carlos, Universidade de S\~ao Paulo, 13560-970 S\~ao Carlos, SP, Brazil}
\author{R. Kaiser} \affiliation{Universit\'e de Nice Sophia Antipolis, CNRS, Institut Non-Lin\'eaire de Nice, UMR 7335, F-06560 Valbonne, France}
\author{R. Bachelard} \affiliation{Instituto de F\'{i}sica de S\~ao Carlos, Universidade de S\~ao Paulo, 13560-970 S\~ao Carlos, SP, Brazil}

\date{\today}
\begin{abstract}
Quasi-resonant scattering of light in two dimensions can be described either as a scalar or as a vectorial electromagnetic wave. Performing a scaling analysis we observe in both cases long lived modes, yet only the scalar case exhibits Anderson localized modes together with extremely long mode lifetimes. We show that the localization length of these modes is influenced only by their position, and not their lifetime. Investigating the reasons for the absence of localization, it appears that both the coupling of several polarizations and the presence of near-field terms are able to prevent long lifetimes and Anderson localization.
\end{abstract}
\maketitle

\section{Introduction}
Multiple scattering of waves has been the subject of intense debates  in the context of disorder-induced Anderson localization~\cite{anderson:1958}. Indeed, since the proposal to use electromagnetic waves in random media instead of electrons in solids exploiting the non-interacting properties of photons at low intensities~\cite{sajeev1987}, many experiments and theoretical studies have been performed. However, despite a decade-long research, the mere existence of Anderson localization of light~\cite{klein:1997,skipetrov2014,skipetrov2015} and its relation to another long predicted phenomenon, namely Dicke super- and subradiance~\cite{dicke1954}, are still not clearly understood~\cite{akkermans2008}. The advent of laser-cooled atoms and their use to study both localization and super- and subradiance motivated the development of ab initio models of interference effects in multiple scattering of light~\cite{piovella2010}. As most experiments are typically performed in a three-dimensional setting, models have also been focused on such 3D configurations. However, both numerical and fundamental aspects of localization strongly depend on the dimension of the explored system~\cite{abraham1979}. For this reason we have focused our efforts on a 2D system, where a precise study of the eigenvalues and eigenmodes of the system is more efficient than in 3D, because larger 'volumes' can be simulated for a given number of scatterers. One further advantage is that the reduced dimensionality allows for a direct comparison of eigenvalues and eigenvectors between two regimes of scattering, one of a scalar model of light, the other of a vectorial model of light where the wave polarization needs to be accounted for. This comparison recently revealed important differences observed for the eigenvalues of the relevant effective Hamiltonians~\cite{skipetrov2014,skipetrov2015,bellando2014}.

\begin{figure}[!ht]
\centering \includegraphics[width=1\linewidth]{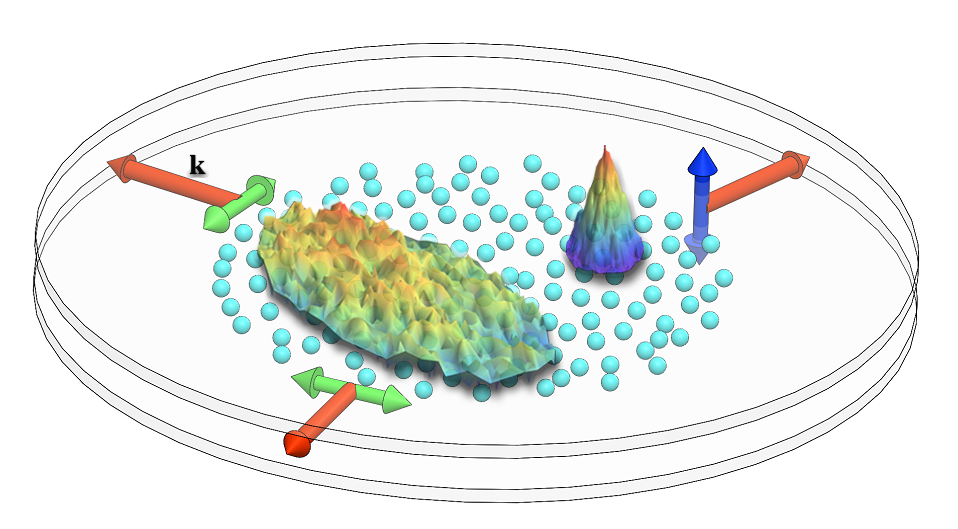}
\caption{\label{fig:setup} (Color online) Two-dimensional scattering scheme: The radiation, of wavenumber $k$ close to the atomic transition $k_a=\omega_a/c\approx k$, is confined in the $(x,y)$ plane, i.e., it has wavevectors of the form $\mathbf{k}=k(\cos\theta,\sin\theta,0)$. 
Two eigenvectors are shown: a localized  $s-$polarized mode in the upper right part and a extended $p-$polarized mode in the left part.}
\end{figure}

In this work, we investigate resonant scattering in a two-dimensional set-up, i.e., the light scattering and propagation are confined to two dimensions. This configuration may be realized, e.g., with a disordered arrangement of scatterers in microwave cavities \cite{mortessagne2007}, in photonic crystals~\cite{busch2007}, near surface plasmons or with laser-cooled atoms located in an off-resonance optical cavity. In this geometry, the polarization orthogonal to the plane, called $s$-polarization, cannot couple  through the scatterers to the planar (or $p$-) polarizations, hence it is described by a scalar light model. The two $p$-polarizations of the electromagnetic waves do couple, and this vectorial-like scattering includes near-field terms (see scheme in Fig.\ref{fig:setup}). Rotating the polarization of an incident wave allows to switch between the scalar or vectorial regime, between the presence or the absence of polarization degrees of freedom and near field terms, making it an ideal tool to investigate the role of polarization in localization and subradiance.

In section II we present a detailed derivation of the linear differential equations  that rules the population evolution of atomic transitions. The spectral properties of these equations are investigated in section III, where scalar and vector scattering are compared via scaling analysis. Finally, we present our conclusions in section IV highlighting uncorrelation between spatial and temporal localization of light.

\section{Model}
Two-dimensional light scattering was investigated in microwave cavities, where light with a polarization orthogonal to the plates approximately obeys Helmholtz 2D scalar equation, and Anderson localization was observed~\cite{mortessagne2007}. Another possibility to emulate 2D light scattering is a cloud of cold atoms with no Doppler broadening located inside an off-resonant optical cavity made by two metallic disks whose diameter is much larger than their mutual distance. This system, which is closer to situations previously studied in 3D, constitutes the toy model system we will explore in this paper. Let us consider an homogeneous disk-shaped cloud of $N$ motionless atoms sitting at randomly distributed positions $\mathbf{r}_j=(x_j,y_j,z_j)$ with $j=1,\cdots,N$, for which non-radiative interactions are neglected. Instead, only virtual and real photons couple the atoms within the optical cavity (axis $z$) whose resonance frequency is significantly detuned from the atomic transition $\omega_a$. The electric dipole transitions occur between one non-degenerate ground state $\ket{g_j}$, related to angular momentum $\ell=0$ and a triply degenerate excited state $\ket{e_j^m}$, where $m=0,\pm1$ indicate the projections of the angular momentum $\ell=1$ over the quantization axis $z$. We consider 2D scattering restricted to a radial direction in the $(x,y)$ plane with a surface density of the atomic cloud $\rho=N/\pi R^2$, where $R$ is the cloud radius.

The interaction of the atoms with the radiation field is given by the following Hamiltonian:
\begin{eqnarray}
\hat{H}&=&\frac{\hbar\omega_{a}}{2}\sum_{j=1}^{N}\hat{\sigma}_{z,j}+\hbar\sum_{\mathbf{k},s}\omega_{k}\left(\hat{a}_{\mathbf{k},s}^{\dagger}\hat{a}_{\mathbf{k},s}+\frac{1}{2}\right)\nonumber
\\ &&+\hbar\sum_{j=1}^{N}\sum_{m=-1}^{1}\sum_{\mathbf{k},s}\left(\hat{a}_{\mathbf{k},s}^{\dagger}e^{-i\mathbf{k}\cdot\mathbf{r}{}_{j}}+\hat{a}_{\mathbf{k},s}e^{i\mathbf{k}\cdot\mathbf{r}{}_{j}}\right)\times\nonumber
\\ &&\times\left(g_{\mathbf{k},s}^{j,m}\hat{\sigma}_{j}^{\left(m\right)}+g_{\mathbf{k},s}^{j,m*}\hat{\sigma}_{j}^{\dagger\left(m\right)}\right),\label{eq:hint}
\end{eqnarray}
with $\omega_a$ the atomic transition frequency, $\hat{\sigma}_{z,j}$ the atomic diagonal term, whereas $\hat{\sigma}_{j}^{\left(m\right)}=\left|g_{j}\right\rangle \left\langle e_{j,m}\right|$ and $\hat{\sigma}_{j}^{\dagger\left(m\right)}=\left|e_{j,m}\right\rangle \left\langle g_{j}\right|$ are the lowering and lifting atomic operators. $\hat{a}_{\mathbf{k},s}^{\dagger}$ and $\hat{a}_{\mathbf{k},s}$ refer to the creation and annihilation of a photon for mode $\mathbf{k}$, with frequency $\omega_k$. The coupling coefficient reads $g_{\mathbf{k},s}^{j,m}=\hat{\mathbf{e}}_{\mathbf{k},s}\cdot\mathbf{d}_{j,m}\sqrt{\omega_{\mathbf{k}}/2\hbar\epsilon_{0}V}$, with $V$ the quantization volume and $\mathbf{d}_{j,m}=\left\langle g_{j}\right|e\mathbf{r}_{j}\left|e_{j,m}\right\rangle$ the dipole matrix element (with $e$ is the electron charge). Thus, the two first terms in \eqref{eq:hint} are the free energy contribution and the last term corresponds to the interaction with the vacuum modes. 

We then use the commutation relations
\begin{subequations}
\begin{eqnarray}
\left[\hat{\sigma}_{j}^{\dagger\left(m\right)},\hat{\sigma}_{j'}^{\left(m'\right)}\right]&\approx& -\mathds{1}_{j'}^{m'}\delta_{j,j'}\delta_{m,m'},\label{eq:unity}
\\ \left[\hat{a}_{\mathbf{k},s},\hat{a}_{\mathbf{k}',s'}^{\dagger}\right] &=& \delta_{s,s'}\delta_{\mathbf{k},\mathbf{k}'},
\\ \left[\hat{\sigma}_{j}^{\left(m\right)},\hat{\sigma}_{z,j'}^{\left(m'\right)}\right] &=& 2\hat{\sigma}_{j'}^{\left(m'\right)}\delta_{j,j'}\delta_{m,m'},
\end{eqnarray}
\end{subequations}
where the approximation in the first of these equations correspond to the linear optics regime $\left|e_{j,m}\right\rangle \left\langle e_{j,m}\right|-\left|g_{j}\right\rangle \left\langle g_{j}\right|\approx -\mathds{1}_{j}^{m}$ (we eliminate the possibility of multiexcitation in the system), to obtain the Heisenberg equations for the operators:
\begin{subequations}
\begin{eqnarray}
\frac{d\hat{\sigma}_{j}^{\left(m\right)}}{dt}+i\omega_{a}\hat{\sigma}_{j}^{\left(m\right)}&=&-i\sum_{\mathbf{k},s}g_{\mathbf{k},s}^{j,m*}\left(\hat{a}_{\mathbf{k},s}e^{i\mathbf{k}\cdot\mathbf{r}{}_{j}}+\hat{a}_{\mathbf{k},s}^{\dagger}e^{-i\mathbf{k}\cdot\mathbf{r}{}_{j}}\right),\label{eq:atoms}
\\ \frac{d\hat{a}_{\mathbf{k},s}}{dt}+i\omega_{\mathbf{k}}\hat{a}_{\mathbf{k},s}&=&-i\sum_{j=1}^{N}\sum_{m=-1}^{1}\left(g_{\mathbf{k},s}^{j,m}\hat{\sigma}_{j}^{\left(m\right)}+g_{\mathbf{k},s}^{j,m*}\hat{\sigma}_{j}^{\dagger\left(m\right)}\right)e^{-i\mathbf{k}\cdot\mathbf{r}{}_{j}}.\label{eq:photons}
\end{eqnarray}
\end{subequations}
Equations \eqref{eq:atoms} and \eqref{eq:photons} show a correlated dynamics between atomic levels and vacuum modes. 

The radiation field plays the role of a reservoir for atoms and is composed of an infinite number of degrees of freedom $\mathbf{k},s$, so it is convenient to trace over these. Using the unitary transformations $\hat{\sigma}_{j}^{\left(m\right)}\rightarrow\hat{\sigma}_{j}^{\left(m\right)}e^{i\omega_{a}t}$ and $\hat{a}_{\mathbf{k},s}\rightarrow\hat{a}_{\mathbf{k},s}e^{i\omega_{\mathbf{k}}t}$, we obtain the reduced equation evolution for the atomic open system
\begin{widetext}
\begin{eqnarray}
\frac{d\hat{\sigma}_{j}^{\left(m\right)}}{dt}&=&-\sum_{l,n}\sum_{\mathbf{k},s}g_{\mathbf{k},s}^{j,m*}\int_{0}^{t}d\tau\left(g_{\mathbf{k},s}^{l,n}\hat{\sigma}_{l}^{\left(n\right)}\left(t-\tau\right)e^{i\omega_{a}\tau}+g_{\mathbf{k},s}^{l,n*}\hat{\sigma}_{l}^{\dagger\left(n\right)}\left(t-\tau\right)e^{i\omega_{a}\left(2t-\tau\right)}\right)\left(e^{i\omega_{\mathbf{k}}\tau+i\mathbf{k}\cdot\mathbf{r}_{jl}}-c.c\right).
\end{eqnarray}
\end{widetext}
We can apply the rotating wave approximation and neglect the fast oscillating terms proportional to $e^{2i\omega_{a}t}$. Assuming that the photon transit time inside the atomic cloud is much shorter than the emission decay time, we can perform the Markov approximation $\hat{\sigma}_{l}^{\left(n\right)}\left(t-\tau\right)\approx\hat{\sigma}_{l}^{\left(n\right)}\left(t\right)$ so the atomic transitions evolves according to the closed set of equations
\begin{eqnarray}
\frac{d\hat{\sigma}_{j}^{\left(m\right)}}{dt}=-\frac{1}{2}\sum_{n=-1}^{1}\sum_{l=1}^{N}K_{m,n}\left(\mathbf{r}_{jl}\right)\hat{\sigma}_{l}^{\left(n\right)},\label{eq:master}
\end{eqnarray}
where the scattering kernel is defined as
\begin{eqnarray}
K_{m,n}\left(\mathbf{r}_{jl}\right)\equiv2\int_{0}^{\infty}d\tau e^{i\omega_{a}\tau}\sum_{\mathbf{k},s}g_{\mathbf{k},s}^{j,m*}g_{\mathbf{k},s}^{l,n}\left(e^{-i\omega_{\mathbf{k}}\tau+i\mathbf{k}\cdot\mathbf{r}{}_{jl}}-c.c.\right),\label{eq:kernelg}
\end{eqnarray}
with $\mathbf{r}_{jl}=\mathbf{r}_{j}-\mathbf{r}_{l}$, and where the upper limit in the above integral has been extrapolated to $t\rightarrow\infty$ according to the Markov approximation. The spontaneous emission processes is naturally three-dimensional and the above scattering kernel a priori contains all light modes in 3D space, so all transitions may be coupled. Under the assumption of an effective two-dimensional scattering of light, we can perform in the continuous 3D density of modes the following approximation:
\begin{equation}
\sum_{\mathbf{k}}\rightarrow\frac{V}{\left(2\pi\right)^{3}}\int d^{3}\mathbf{k}\delta\left(\theta-\pi/2\right),\label{eq:condition2d}
\end{equation}
where $\theta$ corresponds to the azimuthal angle in spherical coordinates, and $V$ now refers to the quantization volume delimited by the cavity. Observe we are treating the radiation field inside a volume, however the density of modes is practically parametrized in polar coordinates as $\mathbf{k}=k\left(\cos\phi,\sin\phi,0\right)$. The usual relation for polarization vectors $\sum_{s}\hat{\mathbf{e}}_{\mathbf{k},s}^{\mu}\hat{\mathbf{e}}_{\mathbf{k},s}^{\nu}=\delta_{\mu,\nu}-\hat{\mathbf{k}}_{\mu}\hat{\mathbf{k}}_{\nu}$ ($\mu$, $\nu$ the Cartesian components) here turns into
\begin{eqnarray}
\sum_{s}g_{\mathbf{k},s}^{j,m*}g_{\mathbf{k},s}^{l,n}=
\frac{kc}{2\hbar\epsilon_{0}V}\left[\mathbf{d}_{j,m}^{z*}\mathbf{d}_{l,n}^{z}+\sum_{\mu,\nu\neq z}\mathbf{d}_{j,m}^{\mu*}\mathbf{d}_{l,n}^{\nu}\left(\delta_{\mu,\nu}-\hat{\mathbf{k}}_{\mu}\hat{\mathbf{k}}_{\nu}\right)\right].
\end{eqnarray}
Therefore, the 2D condition \eqref{eq:condition2d} decouples the component $\mathbf{d}_{j,m}^{z}$ of the dipole matrix elements from  $\mathbf{d}_{j,m}^{x}$ and $\mathbf{d}_{j,m}^{y}$. This phenomena is absent in 3D scattering, where all components of $\mathbf{d}_{j,m}$ are coupled, whereas at the other end, 1D case exhibits all components trivially uncoupled.

The single atom decay into two-dimensional vacuum modes is given by
\begin{widetext}
\begin{eqnarray}
K_{m,n}^{2D}\left(\mathbf{r}_{jl}=0\right)&=&\mathbf{d}_{j,m}^{z*}\mathbf{d}_{l,n}^{z}\int_{0}^{\infty}\frac{ck^{3}dk}{\left(2\pi\right)^{3}\hbar\epsilon_{0}}\int_{0}^{2\pi}d\phi\int_{0}^{\infty}d\tau e^{i\omega_{a}\tau}\left(e^{-i\omega_{\mathbf{k}}\tau}-e^{i\omega_{\mathbf{k}}\tau}\right)\nonumber
\\ &&+\sum_{\mu,\nu\neq z}\mathbf{d}_{j,m}^{\mu*}\mathbf{d}_{l,n}^{\nu}\int_{0}^{\infty}\frac{ck^{3}dk}{\left(2\pi\right)^{3}\hbar\epsilon_{0}}\int_{0}^{2\pi}d\phi\left(\delta_{\mu,\nu}-\hat{\mathbf{k}}_{\mu}\hat{\mathbf{k}}_{\nu}\right)\int_{0}^{\infty}d\tau e^{i\omega_{a}\tau}\left(e^{-i\omega_{\mathbf{k}}\tau}-e^{i\omega_{\mathbf{k}}\tau}\right),\label{eq:kzero}
\end{eqnarray}
\end{widetext}
where the index $2D$ applied on $K_{m,n}^{2D}$ means the general kernel particularized to the two-dimensional scattering. The time integral present in equation \eqref{eq:kzero} solves by using the relation
\begin{eqnarray}
\int_{0}^{\infty}d\tau e^{i\left(\omega_{a}\pm\omega_{k}\right)\tau}=\frac{\pi}{c}\delta\left(k\pm k_{a}\right)\pm iP\frac{1}{\omega_{a}\pm\omega_{k}},\label{eq:cauchy}
\end{eqnarray}
where $P$ refers to the Cauchy principal value. This term gives rise to the Lamb shift, a single atom energy shift which is due to its interaction with the radiation field. We will here neglect it, as it simply corresponds to a renormalization of the energy; remark that we do not neglect the so-called {\it collective} Lamb shift, that rises from the interaction between the atoms via virtual photons, as it is still present in the final scattering kernel. 

Using the relations
\begin{subequations}
\begin{eqnarray}
\mathbf{d}_{j,m}=d\left(i\frac{\left|m\right|}{\sqrt{2}},\frac{m}{\sqrt{2}},1-\left|m\right|\right),
\\ \int_{0}^{2\pi}d\phi\left(\delta_{m,n}-\hat{\mathbf{k}}_{\mu}\hat{\mathbf{k}}_{\nu}\right)=\pi\delta_{m,n},\label{eq:dipole}
\end{eqnarray}
\end{subequations}
where the expression for $\mathbf{d}_{j,m}$ in \eqref{eq:dipole} includes choosing the quantization axis over $z$, and plugging \eqref{eq:cauchy} into \eqref{eq:kzero}, we obtain the two different decay rates
\begin{subequations}
\begin{eqnarray}
\Gamma_0&=&K_{0,0}\left(\mathbf{r}_{jl}=0\right)=\frac{k_{a}^{3}d^{2}}{4\pi\hbar\epsilon_{0}},\label{eq:rates2}
\\ \Gamma_1&=&K_{\pm1,\pm1}\left(\mathbf{r}_{jl}=0\right)=\frac{k_{a}^{3}d^{2}}{8\pi\hbar\epsilon_{0}}=\frac{\Gamma_0}{2}.\label{eq:rates}
\end{eqnarray}
\end{subequations}
By looking at \eqref{eq:cauchy}, one can see only the term $k=k_a$ contributes. Therefore, the uncoupling of $\mathbf{d}_{j,m}^{z}$ from the other components of $\mathbf{d}_{j,m}$ causes an anisotropy on spontaneous emission process, since the lifetime of transition $e_{j,m=0}\rightarrow g_{j}$ is twice shorter than $e_{j,m=\pm1}\rightarrow g_{j}$. These decay rates will predict the coexistence of two scattering subsystems with different time scales, which  vector nature of light will be crucial to select each subsystem is active.
 
Finally, we address the collective term by calculating the integrals describing the coupling between the atoms via the radiation field
\begin{widetext}
\begin{eqnarray}
K_{m,n}^{2D}\left(\mathbf{r}{}_{jl}\neq0\right)&=&\mathbf{d}_{j,m}^{z*}\mathbf{d}_{l,n}^{z}\int_{0}^{\infty}d\tau e^{i\omega_{a}\tau}\int_{0}^{\infty}\frac{ck^{3}dk}{\left(2\pi\right)^{3}\hbar\epsilon_{0}}\int_{0}^{2\pi}d\phi\left(e^{-i\omega_{\mathbf{k}}\tau+i\mathbf{k}\cdot\mathbf{r}{}_{jl}}-c.c\right)\nonumber
\\&&+\sum_{\mu,\nu\neq z}\mathbf{d}_{j,m}^{\mu*}\mathbf{d}_{l,n}^{\nu}\int_{0}^{\infty}d\tau e^{i\omega_{a}\tau}\int_{0}^{\infty}\frac{ck^{3}dk}{\left(2\pi\right)^{3}\hbar\epsilon_{0}}\int_{0}^{2\pi}d\phi\left(\delta_{\mu,\nu}-\hat{\mathbf{k}}_{\mu}\hat{\mathbf{k}}_{\nu}\right)\left(e^{-i\omega_{\mathbf{k}}\tau+i\mathbf{k}\cdot\mathbf{r}{}_{jl}}-c.c\right).\label{eq:kdiff}
\end{eqnarray}
\end{widetext}
Since the density of modes is non-zero only in the $(x,y)$ plane, the $z$ component of the atoms positions does not come into play, so in the relation
\begin{eqnarray}
\hat{\mathbf{k}}_{\mu}\hat{\mathbf{k}}_{\nu}e^{\pm i\mathbf{k}\cdot\mathbf{r}_{jl}}=-\frac{\partial^{2}}{\partial x_{jl}^{\mu}\partial x_{jl}^{\nu}}e^{\pm i\mathbf{k}\cdot\mathbf{r}_{jl}},\label{eq:trick}
\end{eqnarray}
where $x_{jl}^{\mu}$ actually spans only $(x_{jl},y_{jl})$. The angular integral then reads
\begin{eqnarray}
\int_{0}^{2\pi} e^{\pm ikx_{jl}\cos\phi\pm iky_{jl}\sin\phi}d\phi=2\pi J_{0}\left(kr_{jl}\right),\label{eq:angularintegral}
\end{eqnarray}
where $J_0$ denotes the Bessel function of the first kind and of order $0$, and $r_{jl}=\sqrt{x_{jl}^{2}+y_{jl}^{2}}$ the Euclidean distance between each pair of atoms in the plane. Despite the integrands in \eqref{eq:kdiff} diverge in the limit $k\rightarrow\infty$, the modulus of the wavevectors $k$ vary only slightly around $k=k_{a}$ (quasi-elastic scattering). We then apply the Wigner and Weisskopf approximation which approximates powers of $k$ in the integral as $k_{a}$. Using the relation
\begin{eqnarray}
\int_{0}^{\infty}d\tau e^{i\omega_{a}\tau}\int_{0}^{\infty}dkJ_{0}\left(kr_{jl}\right)\sin\left(kc\tau\right)=\frac{\pi i}{2c}H_{0}\left(k_{a}r_{jl}\right),\label{eq:kintegral2}
\end{eqnarray}
where $H_{\alpha}$ is the Hankel function of the first kind and of order $\alpha$, we can calculate $K_{m,n}^{2D}\left(\mathbf{r}{}_{jl}\neq0\right)$ from the action of the second order derivative of $H_{0}\left(k_{a}r_{jl}\right)$ with respect to $x_{jl}^{\mu}$. Practically, we get
\begin{subequations}
\begin{eqnarray}
K_{0,0}^{2D}\left(\mathbf{r}_{jl}\neq0\right)&=&\Gamma_0 H_0\left(k_{a}r_{jl}\right),
\\ K_{\pm1,\pm1}^{2D}\left(\mathbf{r}_{jl}\neq0\right)&=&\Gamma_1 H_0\left(k_{a}r_{jl}\right),
\\ K_{\pm1,\mp1}^{2D}\left(\mathbf{r}_{jl}\neq0\right)&=&\Gamma_1 H_2\left(k_{a}r_{jl}\right)e^{\pm 2i\varphi_{jl}}.
\label{eq:kintegral}
\end{eqnarray}
\end{subequations}
These coefficients allow to obtain the following set of equations for the atomic operators $\hat{\sigma}_{j}^{\left(m\right)}$:
\begin{widetext}
\begin{subequations}
\begin{eqnarray}
\frac{d\hat{\sigma}_{j}^{\left(0\right)}}{dt}=-\frac{\Gamma_{0}}{2}\hat{\sigma}_{l}^{\left(0\right)}-\frac{\Gamma_{0}}{2}\sum_{l=1}^{N}H_{0}\left(kr_{jl}\right)\hat{\sigma}_{l}^{\left(0\right)}, \label{eq:gs}
\\ \frac{d\hat{\sigma}_{j}^{(\pm1)}}{dt}=-\frac{\Gamma_{1}}{2}\hat{\sigma}_{j}^{(\pm1)}-\frac{\Gamma_{1}}{2}\sum_{l\neq j}\left(H_{0}\left(kr_{jl}\right)\hat{\sigma}_{l}^{(\pm1)}+e^{2i\varphi_{jl}}H_{2}\left(kr_{jl}\right)\hat{\sigma}_{l}^{(\mp1)}\right),\label{eq:vector}
\end{eqnarray} 
\end{subequations}
\end{widetext}
where $\tan\varphi_{jl}=(y_{j}-y_{l})/(x_{j}-x_{l})$. In both equations \eqref{eq:gs} and \eqref{eq:vector}, the atoms are coupled together through the same sublevel with a kernel term $H_0(kr)$ that scales as a 2D spherical wave $e^{ikr}/\sqrt{r}$ at large distances. This $1/\sqrt{r}$ scaling in 2D corresponds to long range coupling so we expect global coupling (or cooperative effects) to dominate over nearest neighbor coupling. In the vectorial case, the $m=\pm1$ sublevels are additionally coupled via a $H_2$ term which also scales as $e^{ikr}/\sqrt{r}$ at long range. However, with respect to global versus local interactions, $H_0$ diverges at the origin only as $\log(kr)$, so that the contribution of a small volume around the particle is finite ($\int_0^{r_0} \log(kr)2\pi r dr<\infty$); Instead, the $H_2$ term diverges at the origin as $1/r^2$, which means that the interaction between the $\pm 1$ sublevels is dominated by the close neighbors/near-field terms at high densities ($\int_{r_-}^{r_0} 2\pi r dr/r^2\underset{r_- \to 0}{\sim}-2\pi \log(r_-)$). 

Differently from the 3D case where scalar light is only an approximation for dilute systems, and where all sublevels are normally coupled, in 2D geometries scalar model holds for high densities. Yet, controlling the polarization of the injected light allows to select either purely scalar or vectorial properties which make our approach quite versatile. In the end, two decoupled scattering subsystems appear: one involving a single sublevel of the excited state (the scalar case), the other one involving the remaining two sublevels (the vectorial case). As one can note the scalar and vector kernels are not decoupled through energy shifts like 3D work in Ref.\cite{skipetrov2015}, but rather by geometrical constraints. The microwave or optical cavity reshapes the density of electromagnetic modes into two dimensions.

The scattered field at a point $\mathbf{r}=\left(x,y\right)$ is calculated by a superposition of annihilation operators, namely
\begin{eqnarray}
\hat{\mathbf{E}}\left(\boldsymbol{\mathbf{r}}\right)=\sum_{\mathbf{k},s}\varepsilon_{\mathbf{k}}\hat{\mathbf{e}}_{\mathbf{k},s}a_{\mathbf{k},s}e^{-i\omega_{\mathbf{k}}t+i\mathbf{k}\cdot\mathbf{r}},
\end{eqnarray}
where $\varepsilon_{\mathbf{k}}=\sqrt{\hbar\omega_{\mathbf{k}}/2\epsilon_{0}V}$. With  similar procedures used up to here, using \eqref{eq:photons}, it can be shown to lead to the following classical equation
\begin{eqnarray}
\hat{\mathbf{E}}\left(\boldsymbol{\mathbf{r}}\right)&=&\hat{\mathbf{z}}\frac{\hbar\Gamma_{0}}{2di}\sum_{l=1}^{N}\left(H_{0}\left(k_{a}\left|\mathbf{r}-\mathbf{r}_{l}\right|\right)+\frac{i}{\pi k_{a}\left|\mathbf{r}-\mathbf{r}_{l}\right|}\right)\hat{\sigma}_{l}^{\left(0\right)}\nonumber
\\ &&+\frac{\hbar\Gamma_{1}}{2d}\sum_{l=1}^{N}\sum_{m=\pm1}\mathbf{e}_{m}e^{-2mi\varphi_{l}}H_{2}\left(k_{a}\left|\mathbf{r}-\mathbf{r}_{l}\right|\right)\hat{\sigma}_{l}^{\left(m\right)},\nonumber
\\ &&+\frac{\hbar\Gamma_{1}}{2d}\sum_{l=1}^{N}\sum_{m=\pm1}\mathbf{e}_{m}H_{0}\left(k_{a}\left|\mathbf{r}-\mathbf{r}_{l}\right|\right)\hat{\sigma}_{l}^{\left(-m\right)}
 \end{eqnarray}
with $\tan\varphi_l=(y-y_{l})/(x-x_{l})$ and $\mathbf{e}_{\pm}=\left(\hat{\mathbf{x}}\pm i\hat{\mathbf{y}}\right)/\sqrt{2}$. The $m=0$ sublevel is coupled only to light with polarization along $z$, whereas the $m=\pm1$ sublevels are coupled together through $p$-polarizations. 

\section{Spectral Analysis of the Linear Equations}
We now turn our attention to the spectral properties of the system. The scattering modes are the eigenmodes $\Psi^{(n)}$ of the linear equations (\ref{eq:gs}) and (\ref{eq:vector}), where $n$ labels the eigenmodes. Their lifetime $1/\gamma_n$ and energy $\omega_n$ are given by the real and imaginary part of the associated eigenvalues, respectively. Defining $\psi_j^{(n)}=(\Psi^{(n)})_j$, the modes can also be characterized by their inverse participation ratio (IPR) $(\sum_{j}|\psi_j^{(n)}|^{4})/(\sum_{j}|\psi_j^{(n)}|^{2})^2$, that quantifies the (inverse) number of atoms substantially involved to the scattering mode. In the vectorial model we renormalize the IPR to remain 1/2 for pairs.

\begin{figure} [!ht]
\centering
\includegraphics[width=1\linewidth]{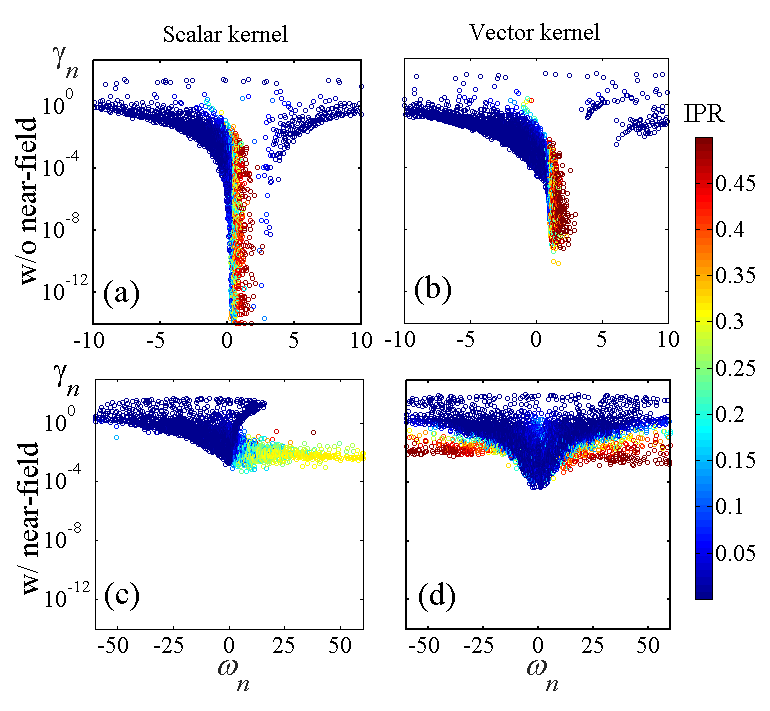}
\caption{\label{fig:panel} (Color online) Inverse participation ratio of the scattering modes in the complex plane of the eigenvalues $(\gamma_n,\omega_n)$ for scalar light (a) without and (c) with near-field terms, in units of $\Gamma_0$, and vectorial light (b) without and (d) with near-field terms, in units of $\Gamma_1$, (a) and (d) being the physical cases. Simulations realized for an homogeneous disk cloud of $N=5000$ particles with an homogeneous $\rho/k^{2}=1$ density.}
\end{figure}

\begin{figure} [!ht]
\centering \includegraphics[width=1\linewidth]{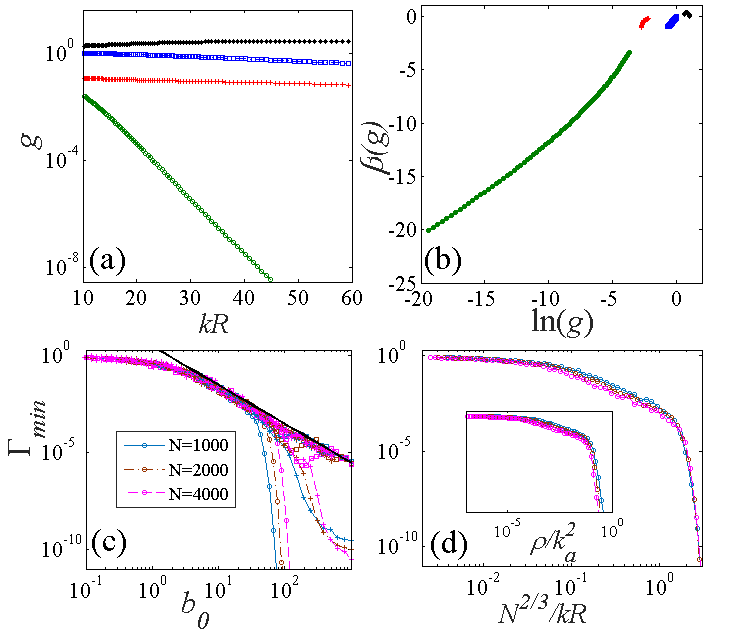}
\caption{\label{fig:g_beta} (Color online) (a) Thouless number $g$ as a function of the dimensionless system size $kR$, for density $\rho/k^2=0.3$ (b) scaling function $\beta$. Longest mode lifetimes  for different number of particles (c) as a function of the optical thickness $b_0$ and for all four scattering models and (d) as a function of the dimensionless parameter $N^{2/3}/kR$ for scalar light. The circles are for scalar light, the squares for vectorial light, the stars for scalar light {\it with near field} and the plus for vectorial light {\it without near field}. The black line in (c) stands for the $\sim 1/b_0^2$ curve of the radiation trapping lifetime. In (c) and (d), blue, brown and magenta curves correspond respectively to simulations with $N=1000$, $2000$ and $4000$ particles. The insert in (d) shows $\Gamma_{min}$ as a function of atomic density, indicating that density is not the best scaling function for $\Gamma_{min}$, even in the high density limit.}
\end{figure}

In the scalar case, the eigenvalue distribution shown in Fig.\ref{fig:panel}(a) exhibits strongly subradiant modes, which we define as modes with very long lifetimes ($\gamma_n\ll\Gamma_0$). The distribution can be used to look for a single parameter scaling, by computing a spectral overlap function conveniently defined as $g=\langle 1/\gamma_n\rangle^{-1}/\langle \omega_n-\omega_{n-1}\rangle$, where the modes $n$ are ordered by increasing energy. In line with the 3D results, we observe a monotonic decrease of $g$ with the system size for scalar light (see Fig.\ref{fig:g_beta}(a)). Consequently, the scaling function $\beta=\partial \ln g/\partial\ln (kR)$ is clearly negative for all values of $g$ (see Fig.\ref{fig:g_beta}(b)), as expected for Anderson localization in 2D. We note that this function $g$ is only one among several possibilities of defining a spectral overlap and has not been shown to be unequivocally related to transport properties of electromagnetic radiation.

The dimensionless scaling parameter g is defined in the scaling theory \cite{edwards1972} as the ratio between the Heisenberg time and the Thouless time. The former corresponds to the time associated to the mean spacing between the energy levels, i.e. $\hbar/\langle E_n-E_{n-1}\rangle$, which in our case reads $1/\langle \omega_n-\omega_{n-1}\rangle$. The latter corresponds to the time necessary to a photon to escape from the sample, and in our open system with eigenmodes of lifetime $1/\gamma_n$, we define it as $\langle 1/\gamma_n\rangle$. Following this interpretation, the localization regime is characterized by a Thouless (diffusion) time that becomes larger than the Heisenberg one. 

In the vectorial case exhibited in Fig.\ref{fig:panel}(d) we observe a dramatically different eigenvalue distribution. Indeed, even though long-lived modes ($\gamma_n\ll\Gamma_0$) exist, they are limited to values larger by several orders of magnitude compared to the scalar case. The corresponding spectral overlap also shows a distinct behavior, with $g$ almost independent of the system size $kR$ [see Fig.\ref{fig:g_beta}(a)], yielding a scaling function $\beta$ close to zero, albeit slightly negative [see Fig.\ref{fig:g_beta}(b)]. This behavior of the scaling function $\beta$ might make this vectorial case very interesting to study fine corrections of the atom-atom interactions as it seems to be close to the critical regime.

The above discussion is consistent with the conclusions drawn from the study of eigenvalues in 3D \cite{skipetrov2014, bellando2014}. With the aim to pin down the essential ingredient of the difference between the scalar and vectorial model, we artificially introduced or removed short range terms in the two configurations. More specifically, we removed the near-field coupling from vectorial scattering by substituting $H_2(kr)$ by $H_2(kr)+4i/\pi(kr)^2$, thus suppressing the near fields. The corresponding eigenvalue distribution is shown in  Fig.\ref{fig:panel}(b). Despite the fact that the $\pm1$ sublevels remain coupled, the eigenvalue distribution of the vectorial case without near field terms closely resembles that of the scalar case, even though the smallest values of $\gamma_n$ do not reach the lowest limits obtained in the scalar case. Conversely, if we add a near field term, which we choose as the one present in vectorial scattering $-4i/\pi(kr)^2$, to the scalar kernel $H_0(kr)$, the long lived modes of the purely scalar case disappear [see Fig.\ref{fig:panel}(c)]. The scaling analysis, as well as a thorough analysis of the spatial extension of the modes, confirm that Anderson localization is absent from these altered interactions. 

Focusing on lifetimes, the study of the longest of them $\Gamma_{min}$ first reveals that for low densities ($\rho<0.3$), long lifetimes are caused by the radiation trapping: $\sim 1/b_0^2$~\cite{holstein1947}, with $b_0$ the cloud optical depth~\cite{note} (see Fig.\ref{fig:g_beta}(c)). However, for scalar light, the appearance of the localized modes  for $\rho/k^2>0.3$ comes along with lifetimes much larger than those predicted by radiation trapping, see Fig.\ref{fig:g_beta}(c). These lifetimes are not simply a function of the density $\rho/k^2$ but appear to scale as $N^{2/3}/kR$ and to decay exponentially fast (see Fig.\ref{fig:g_beta}(d)). This result is clearly beyond the standard Anderson localization, where quantities scale as $N/R^2$, or
cooperative effects where it scales as N/R, and calls for new
approaches. Finally, while scalar light {\it with near-fields} exhibits lifetimes that always decay as $1/b_0^2$ (it is almost with the radiation trapping black curve in Fig.\ref{fig:g_beta}(c)), both vectorial light {\it with} and {\it without} near-field exhibit lifetimes longer than that of radiation trapping: These come from atom pairs instead of localized modes, as reveals the analysis of the IPR and of the spatial profiles.

These results suggest that both the presence of near field interaction terms~\cite{skipetrov2014} and coupling of different sublevels can break down long lifetimes and localization.  We also found that removing the anisotropy present in vectorial scattering ($e^{\pm 2i\varphi_{jl}}$ in Eqs.[\ref{eq:vector}]) does not restore localized modes.

\begin{figure} [!ht]
\centering
\includegraphics[width=1.0\linewidth]{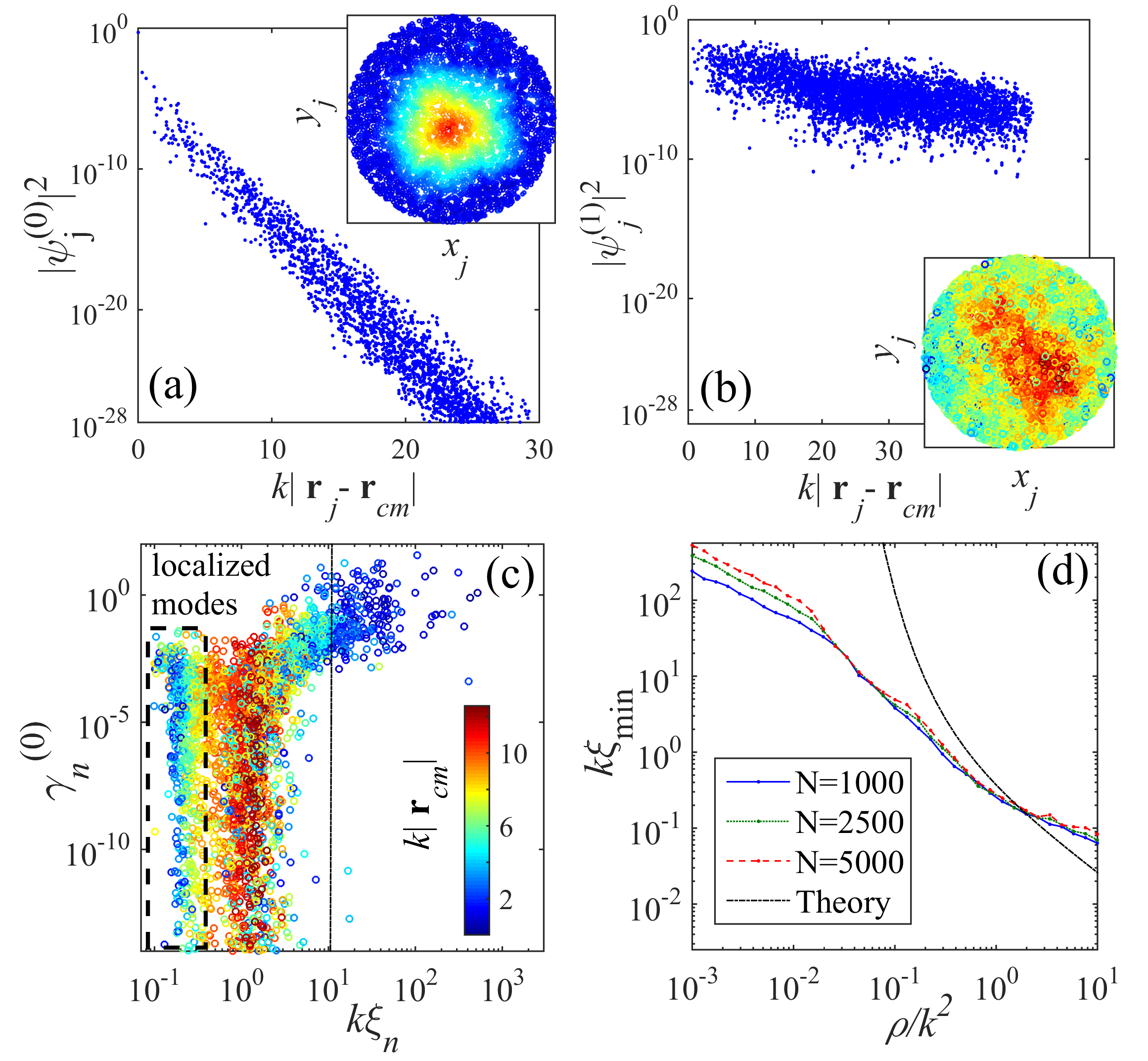}
\caption{\label{fig:profiles} (Color online) Spatial profile of the most subradiant non-pair mode for (a) scalar light (exponentially localized mode) and (b) vectorial light (delocalized mode), as a function of the distance to its center of mass $\mathbf{r}_{cm}$. The 2D profile is exhibited in the inset. (c) Inverse lifetime versus localization length of the modes for scalar light; the localization length $\xi$ is obtained by exponential fit of the spatial profile of the mode ((a) for an example), so it is meaningful only for localized modes, within the dashed-bordered box. More specifically, the strongly subradiant modes lying outside of the box are extended, and so are the superradiant ($\xi$ typically exceeds the system size for these, sign of a full delocalization). (d) Localization length for scalar light as a function of the normalized density, for different scatterer number. The 'theory' line refers to the theoretical prediction $k\xi=(k^2/4\rho)\exp(\pi k^2/8\rho)$. Panels (a-b) were realized for $N=5000$ and $\rho/k^2=1$, so $kR\approx 40$; Panel (c) is for $N=5000$ and $\rho/k^2=10$, so $kR\approx 12.6$, as marked by the dash-dotted line.}
\end{figure}

\begin{figure} [h!]
\centering
\includegraphics[width=1\linewidth]{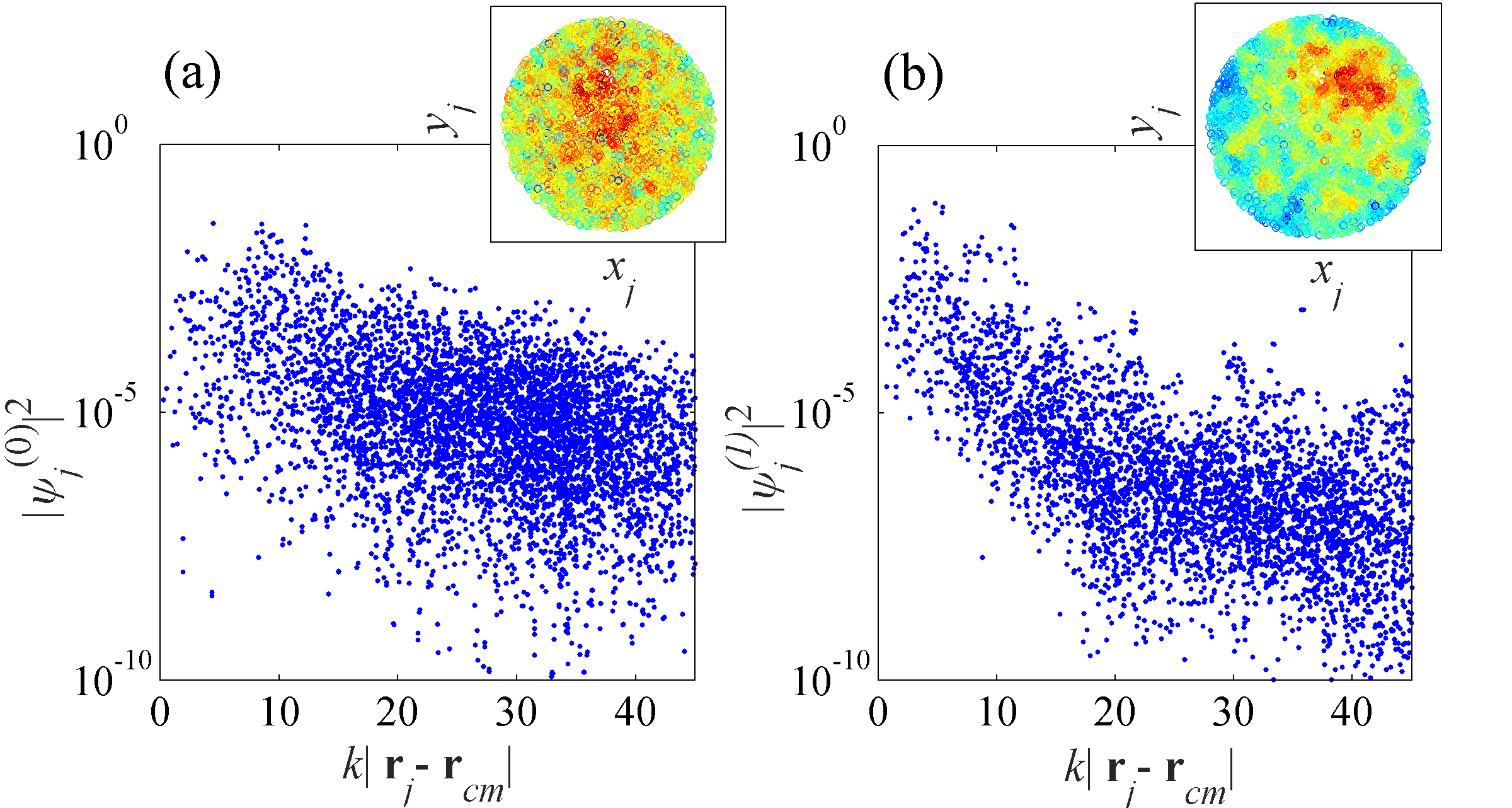}
\caption{\label{fig:alteredprofiles} (Color online) Spatial profile of the most subradiant non-pair mode for (a) scalar light {\it with} near fields and (b) vectorial light {\it without} near fields. Simulations realized for an homogeneous disk cloud of $N=5000$ particles with an homogeneous $\rho/k^{2}=1$ density}
\end{figure}

Our 2D study, apart from the investigation of subradiance and localization in lower dimensions, allows for a more efficient numerical study of the eigenvectors of the dipole-dipole coupling. One aspect of the eigenvector analysis is already seen in Fig.\ref{fig:panel}, where the IPR of the eigenmodes allows, for instance, a clear identification of atomic pairs (red circles in Fig.\ref{fig:panel} corresponding to an IPR close to 0.5, indicating atom pairs). In addition the 2D configuration allows for an easy systematic study of the shapes of the eigenvectors: two typical eigenmodes are shown in Fig.\ref{fig:setup}. The localized mode is spatially well confined [insert in Fig.\ref{fig:profiles}(a)] and has a clear exponential shape over several orders of magnitude [Fig.\ref{fig:profiles}(a)]. Vectorial eigenvectors, on the other hand, are extended over almost the whole system size [insert in Fig.\ref{fig:profiles}(b)], with no indication for an exponential decrease [Fig.\ref{fig:profiles}(b)]. This observation is again in line with previous conclusions in 3D \cite{skipetrov2014, bellando2014}. The scalar light {\it with} near fields and vectorial light {\it without} near fields do not exhibit any exponentially localized modes, but rather extended modes, as can be observed in Fig.~\ref{fig:alteredprofiles}. These scatterings, as well as the vectorial light {\it with} near fields, may however present features of hybrid states where localized and extended subradiant features combine~\cite{celardo2013}.

The localized nature of the strongly subradiant modes is thus confirmed by the analysis of their spatial profile. Furthermore, as long as the mode does not considerably extend over the edge of the atomic cloud, its localization length and lifetime are uncorrelated (uniform filling of the box in Fig.\ref{fig:profiles}c with modes). There is however a correlation between the position of the mode (indicated by the color code, where blue points mark modes at the center of the system) and its localization length. Some modes even mix to surface (whispering gallery) modes that have a much larger spatial extend. Yet there is no correlation between the position of the mode (at the center or near the edge of the cloud) and its lifetime. The absence of correlation between the lifetime of the modes and their localization length calls for a differentiation between spatial and temporal localization. Although all spatially localized modes are subradiant, the shortest localization length may not be associated to the longest lifetimes. This corroborates studies on photon escape rates that failed to observe the localization phase transition \cite{akkermans2008}, and is also highlighted by the fact that spatial localization is affected by boundary effects while temporal localization, surprisingly, does not seem to be. Finally, as can be seen in Fig.\ref{fig:profiles}(d), for densities above $\rho/k^2\sim0.05$ the localization length no longer depends on the system size, but only on the spatial density. These curves are not in agreement with the prediction of localization length from the perturbative approach in the weak disordered regime \cite{patrick1995}.

We have also verified (see Fig.\ref{fig:g_beta}(a)) that the corresponding spectral overlap $g$ and the scaling function $\beta$ for the altered interactions are qualitatively similar to the one of the purely vectorial case, i.e., they deviate only slightly from zero. Similarly, inspection of the eigenvalues did not reveal any spatially localized mode. Together with the above results on lifetimes, this observation suggests that extremely long lifetimes of modes, well beyond radiation trapping ones, come along with spatial localization, i.e., subradiance may be a condition necessary to localization.

\section{Conclusions}
In conclusion, we explored 2D scattering by point scatterers in a scalar and a vectorial limit. Even though our eigenvalue analysis is consistent with previous results and interpretations of localization, our procedure of artificially introducing or removing near field terms combined to a spatial analysis of the eigenfunctions support that very long lifetimes come along with Anderson localization, but both near-field terms and the coupling of polarizations may prevent their emergence. Furthermore, we reported an absence of correlations between lifetime and localization length of localized modes, pointing at the difference between spatial and temporal localization. An important task for the future will be to relate both the 2D and 3D studies to transport properties of electromagnetic waves and to compute observables that can be tested in experiments.

\acknowledgments
We acknowledge financial support from IRSES project COSCALI, from ANR (project LOVE, ANR-14-CE26-0032), from GDRI NSEQO, from CNPq (project PVE 400228/2014-9) and FAPESP. We thank E. Akkermans, D. Delande and S. Skipetrov for stimulating discussions.


\begin{thebibliography}{15}
\bibitem{note} The resonant optical depth is defined as $b_0=\sigma_s\int \rho(0,y)dy$, with $\sigma_0=4/k$, resp. $8/k$, the resonant scattering cross section for scalar, resp. vectorial, light. For an homogeneous disk distribution, one gets $b_0=8N/\pi kR$, resp. $16N/\pi kR$.
\end{thebibliography}
\end{document}